\documentstyle{mn} 

\footnotesize
\newdimen\minuswidth    
\setbox0=\hbox{$-$}
\minuswidth=\wd0
\catcode`@=\active
\def@{\kern\minuswidth}
\newdimen\digitwidth    
\setbox0=\hbox{\rm0}
\digitwidth=\wd0
\catcode`!=\active
\def!{\kern\digitwidth}
\normalsize

\def\etal{et al. }
\def\th{$^{\rm h}$}
\def\tm{$^{\rm m}$}
\def\deg{\ifmmode^{\circ}\else$^{\circ}$\fi} 
\def\min{\ifmmode^{\prime}\else$^{\prime}$\fi} 

\footnotesize
\newdimen\minuswidth    
\setbox0=\hbox{$-$}
\minuswidth=\wd0
\catcode`@=\active
\def@{\kern\minuswidth}
\newdimen\digitwidth    
\setbox0=\hbox{\rm0}
\digitwidth=\wd0
\catcode`!=\active
\def!{\kern\digitwidth}
\normalsize

\title{Millisecond Pulsar Velocities}

\author[M. Toscano et al.]
{ M. Toscano$^{1,2}$, 
J. S. Sandhu$^3$, 
M. Bailes$^2$, 
R. N. Manchester$^{4,1}$, 
M. C. Britton$^2$,
\newauthor 
S. R. Kulkarni$^3$,
S. B. Anderson$^3$,
B. W. Stappers$^5$\thanks{Present Address: Astronomical Institute `Anton Pannekoek', University of Amsterdam,
Kruislaan 403, NL-1098SJ Amsterdam, The Netherlands.}
\\
$^1$Physics Department, University of Melbourne, Parkville, Vic 3052,
Australia.\\ 
$^2$Astrophysics and Supercomputing, Mail No. 31, Swinburne University 
of Technology, PO Box 218, Hawthorn, Vic 3122, Australia.\\
$^3$Department of Astronomy,
Caltech, Pasadena CA 91125. \\
$^4$Australia Telescope National Facility, CSIRO, PO Box
76, Epping, NSW 2121, Australia.\\ 
$^5$Mount Stromlo and Siding Spring Observatories, ANU, Private Bag, 
Weston Creek, ACT 2611, Australia.
}

\begin{document}

\def\lapp{\ifmmode\stackrel{<}{_{\sim}}\else$\stackrel{<}{_{\sim}}$\fi}
\def\gapp{\ifmmode\stackrel{>}{_{\sim}}\else$\stackrel{>}{_{\sim}}$\fi}

\maketitle
\newcommand{\setthebls}{
}
\setthebls

\begin{abstract} 
We present improved timing parameters for 13 millisecond pulsars
(MSPs)  including 9 new proper motion measurements. These new proper
motions  bring to 23 the number of MSPs with measured transverse
velocities.  In light of these new results we present and compare the
kinematic properties of MSPs with those of ordinary pulsars. The mean
transverse velocity  of MSPs was found to be  85$\pm$13 km s$^{-1}$; a
value consistent  with most models for the origin and evolution of
MSPs and approximately a factor of four lower than that of ordinary
pulsars. We also find that, in contrast to young ordinary pulsars, the
vast majority of which are moving away from the Galactic plane,
almost half of the MSPs are moving towards the plane.  This near isotropy
would be expected of a population that has reached dynamic
equilibrium. Accurate measurements  of MSP velocities  have allowed us
to correct their measured spin-down rates for  Doppler  acceleration
effects, and thereby derive their intrinsic magnetic field  strengths
and characteristic ages. We find that close to half of our sample  of
MSPs have a characteristic age comparable to or greater than the age
of the Galaxy.
\end{abstract}


\section{INTRODUCTION}  

The discovery of the first millisecond pulsar (MSP) by Backer et al.
(1982) \nocite{bkh+82} heralded the study of a distinct subclass of
the  pulsar population. The sixty or so MSPs discovered since  then
distinguish themselves from the `ordinary' pulsar population not only
by their shorter rotation periods but also by their greatly reduced
spin-down rates, which imply much weaker magnetic fields and
characteristic  ages comparable to that of the Galactic disc. These
properties, as well as the high incidence of MSPs in binary systems,
suggest that MSPs and ordinary pulsars possess different evolutionary
histories (Alpar \etal 1982; Radhakrishnan \& Srinivasan 1981;
Chanmugam \& Brecher 1987; Bhattacharya \& {van den Heuvel} 1991, and
references therein).  \nocite{acrs82,rs81,cb87,bv91}

MSPs are the paragon of rotational stability and allow us to develop
timing models  that can be used to predict the times of arrival (TOAs)
of their radio pulses, in most cases with an uncertainty of the order
of a few microseconds or less. The millisecond periods that give MSPs
their names  greatly facilitate the process of pulsar timing and
permit very precise  astrophysical measurements to be made
\cite{ktr94}.  Unlike ordinary pulsars which often show the
deleterious effects of `timing noise' over long periods of  time, the
level of such noise in MSPs does not prohibit the measurement of
interesting physical parameters such as proper motion.

Proper motions of radio pulsars have been measured in a variety of
ways. The original method employed conventional optical techniques to
measure the proper motion of the Crab pulsar \cite{tri71}.  However,
this technique has only limited use since $<$1 per cent of pulsars  
are visible at optical wavelengths \cite{wm77,bc88}. The majority of 
pulsar proper motions have been measured using radio interferometric techniques
\cite{las82,bmk+90,fgl+92,hla93,fgm+97}.  A radio-linked
interferometer  is used to measure the separation of the  pulsar and
an extragalactic  reference source in terms of interferometric phase.
Changes in this separation at different epochs  yield the pulsar's
proper motion. A method used to determine pulsar speeds utilises
measured changes  in the pulsar's interstellar scintillation (ISS)
pattern. These changes are interpreted as the result of the pulsar's
motion relative to an hypothetical  scattering screen  midway between
the observer and the pulsar \cite{cor86}. Proper motion measurements
of pulsars based on arrival times were first made by Manchester,
Taylor and Van (1974) \nocite{mtv74} and have been made for a number
of pulsars since \cite{htm77,rtd88,bbm+95}. This technique is most
applicable to pulsars with very low levels of timing noise since
random timing irregularities can exhibit quasi-annual variations that
may be misinterpreted as a positional offset or the pulsar's
motion. The exceptional timing stability of MSPs over periods of
years makes them ideal candidates for proper motion studies.  To date,
proper motions  have been measured for 14 MSPs using arrival-time
methods: PSRs J0437$-$4715  \cite{sbm+97}; J0711$-$6830, J1024$-$0719,
J1744$-$1134, and J2124$-$3358  \cite{bjb+97}; B1257+12 \cite{wol94};
J1713+0747 \cite{cfw94}; B1855+09 and  B1937+21 \cite{ktr94}; B1957+20
\cite{aft94}; J2019+2425 and J2322+2057 \cite{nt95}; J2051$-$0827
\cite{sbm+98} and J2317+1439 \cite{cnt96}. Thirteen MSPs have measured
scintillation speeds: PSRs J0437$-$4715, J0711$-$6830, J1455$-$3330,
J1603$-$7202, J1713+0747,  J1730$-$2304, J1744$-$1134, J1911$-$1114,
J2051$-$0827, J2124$-$3358, J2129$-$5718, J2145$-$0750 and J2317+1439
\cite{jnk98}.

Pulsars are, as a class, high-velocity objects. Lyne and Lorimer
(1994) \nocite{ll94} have suggested that pulsars are born with a mean
velocity  of $\sim$450 km s$^{-1}$. Although Hansen \& Phinney (1997)
\nocite{hp97}  have  derived a lower mean velocity of $\sim$250 $-$
300 km s$^{-1}$, it is clear that pulsar velocities are, on average,
about  a factor of ten higher than those of their progenitor O and B
stars. Two theories have been put forward to explain these high space
velocities: (i) the progenitor (close) binary is disrupted  because
more than half the system mass is lost during the supernova (SN)
explosion \cite{bla61,ggo70},   and (ii) the pulsar receives a  natal
`kick' resulting from the asymmetry of the SN event \cite{shk70}.  The
most likely explanation may well be a combination of both.

MSPs on the other hand have relatively low speeds compared with the
general pulsar population. The known MSP proper motions imply a mean
transverse velocity of $\sim$100 km s$^{-1}$. Although  much work has
been done to model the kinematics of ordinary  pulsars, the lack of
measured MSP proper motions has made constraining MSP velocity models
difficult. Recently Tauris \& Bailes (1996) \nocite{tb96} made Monte
Carlo simulations of a large ensemble of binary systems using a number
of different evolutionary scenarios. The velocity distributions they
derived had means of $\sim$110 km s$^{-1}$ assuming symmetric SNe
explosions and  $\sim$160 km s$^{-1}$ with the inclusion of random
asymmetric kicks. Their models are consistent with observed transverse
velocities. They also predict an anti-correlation between orbital
period and  recoil velocity (although this becomes weaker with the
introduction of random kicks). Cordes \& Chernoff (1997) \nocite{cc97}
have used a parametric likelihood analysis technique to model MSP
dynamics.   The MSPs they evolved were initially given velocities
with an isotropic  angular distribution and magnitudes drawn from  a
Maxwellian distribution. This distribution of speeds had  a dispersion
of $\sigma_{V}$.  Their analysis yield as most likely
$\sigma_{V}$=52$_{-11}^{+17}$  km s$^{-1}$, a three-dimensional
velocity  dispersion of $\sim$84 km s$^{-1}$ and observed transverse
speeds $<$ 200 km s$^{-1}$.

For nearby MSPs, there is another motivation for determining pulsar
velocities.  The transverse motion of nearby MSPs results in an
apparent acceleration that  makes a significant contribution to their
measured spin-down rates \cite{shk70}.  Knowing the transverse
velocity and distances makes it possible to correct for this effect,
extract the intrinsic spin-down rate, and thus calculate improved derived
magnetic field strengths and characteristic ages. Corrected spin-down
rates  ensure that pulsars find their proper place in the
$P$--$\dot{P}$ diagram from which physical conclusions may be drawn.

In \S 2 of this paper we describe our data collection and analysis
methodology. In \S 3 we present improved timing parameters for 13 MSPs
including nine new arrival-time proper motion measurements.  The
measurement of these new proper motions brings the number of MSPs with
known transverse velocities to 23. In light of this, the aim of this
paper is to bring together these results and investigate the kinematic
properties that MSPs share as a class. To this end, in \S 4  we
examine the velocity distribution of MSPs in the context of their
origin and evolution. In particular we compare and contrast  the
velocities and dynamics of MSPs with those of ordinary pulsars.  In \S
4 we also make use of these velocity measurements to derive intrinsic
spin-down rates, magnetic field strengths and characteristic ages. \S
5  summarises our major findings.

\setcounter{table}{0}
\begin{table*}
\begin{minipage}{160mm}
\caption{Timing parameters for 13 MSPs}
\begin{tabular}{lllll}
\hline\hline \\

Pulsar	            
&	PSR J0613$-$0200          
&       PSR J1045$-$4509         
&     	PSR J1455$-$3330        
&     	PSR J1603$-$7202	  
\\ \hline \\				

R. A. (J2000) 	    
& 	06\th 13\tm 43\fs97336(2) 
&	10\th 45\tm 50\fs1927(1) 
&	14\th 55\tm 47\fs9618(3) 
&	16\th 03\tm 35\fs68523(5)  \\	

Decl. (J2000) 	    
&	$-$02\deg 00\min 47\farcs0970(7) 			 
& 	$-$45\deg 09\min 54\farcs195(1)		
& 	$-$33\deg 30\min 46\farcs350(7)				 
& 	$-$72\deg 02\min 32\farcs6517(5)      			  \\

PM in R. A. (mas y$^{-1}$)			  
&	2.0(4)			 
&	$-$5(2)
&	5(6)			 
&	$-$3.5(3)						  \\

PM in Decl. (mas y$^{-1}$)			  
&	$-$7(1)			 
&	6(1)
&    	24(12)			 
&	$-$7.8(5)					          \\

Galactic longitude  
&   	210\fdg41  	     	  
& 	280\fdg85 		 
&	330\fdg72		
& 	316\fdg63  		  \\ 

Galactic latitude   
&	$-$9\fdg30		  
&	12\fdg25	   	 
&	22\fdg56   		
&	$-$14\fdg50  		  \\

Period  (ms) 	    
&   	3.06184403674401(5)	  
&	7.4742241060982(5)	 
&	7.987204795504(3)	
&	14.8419520154668(2)	  \\

Period derivative ($\times 10^{-20}$)  		  
&	0.9572(5)		 
& 	1.766(3)		 
&	2.42(1)			 
&	1.564(2)          	  \\ 	

Epoch (MJD)	    
&	50315.0 		  
&	50277.0			 
& 	50239.0 		
& 	50427.0			  \\

Dispersion Measure (cm$^{-3}$ pc)	          
& 	38.7792(1)		 
&	58.147(5)			
& 	13.578(7)		 
&	38.0501(1)						  \\

Orbital Period (days)  				  
&	1.1985125566(4) 	 
&	4.083529200(6) 		 
&	76.1745675(6) 		 
& 	6.308629570(2) 		 				  \\

a ${\rm sin}i$ (light-s)			  
&	1.0914441(5)		 
&	3.015134(2)			
&  	32.362233(5)		 
&	6.8806582(5)						  \\

Eccentricity	    
&	0.0000038(10)		  
&	0.0000197(13) 		 
&	0.0001697(3) 		
&	0.0000092(1)		  \\

Epoch of periastron (MJD)		          
&	50315.38339(5000)	 
&	50276.26230(6000)		
&	50275.16814(2000)	 
&	50426.11693(2000)					  \\

Longitude of periastron (deg)		          
&	34(14)			 
&	243(5)			
&	223.78(9)		 
&	170.3(9)							  \\

Timing data span (MJD)  			  
& 	49700$-$50931		 
&	49623$-$50931			
&     	49700$-$50779		 
&	49922$-$50933						  \\

RMS timing residual ($\mu$s)			  
& 	3.5			 
&	16.2 				
&	1.4			 
&	1.5							  \\

Number of timing points			  	  
& 	122			 
&	141 				
&	25			 
&	144							  \\
\hline
\end{tabular}
\end{minipage}
\end{table*}

\section{OBSERVATIONS AND ANALYSIS}

We used the Parkes 64-m radio telescope to make regular observations
of  the 17 MSPs discovered in the Parkes 70-cm survey
\cite{mld+96,lml+98}   as well as a number of previously known MSPs
visible from Parkes. The data presented in this paper has been
collected since the end of 1994 using the Caltech correlator
\cite{nav94}.   Over this period, at low frequencies we used a dual
linear polarisation  system centred at 0.66 GHz with a system noise of
90 Jy. Initially, at high frequencies, we used a 1.4 GHz dual circular
polarisation  receiver with a system noise of 40 Jy.   Since 1997
April, the centre receiver of the  Parkes Multibeam receiver system
(dual linear polarisation and 26 Jy system noise) was used in its
place.   At 0.66 GHz  we recorded a single 32 MHz band with 256 lags
per polarisation.  At 1.4 GHz, prior to 1995 July, a single
intermediate-frequency  (IF) band  was recorded with 128 MHz of
bandwidth and 256 lags per polarisation. This system was superseded by
a dual-IF subsystem \cite{san99} that made it possible to record two
128 MHz IFs with 128 lags per polarisation, centred on 1.4 and 1.66
GHz respectively.

The correlator is flexible enough to permit narrow-band observations
with bandwidths of 64, 32, 16, 8 or 4 MHz using all the available
lags. Diffractive interstellar scintillation results in a modulation
of pulse intensity as a function of time and frequency, the time-scale
and bandwidth of which increase with increasing frequency. For some
MSPs, particularly at 0.66 GHz, a few scintles  may be present across
the observing bandwidth. In these cases it is feasible to centre an
appropriately-sized narrow recording band over the scintillation
maximum. The increased frequency resolution reduces dispersion
smearing, and since most of the pulsar's flux density is concentrated
in the scintles, there is a significant gain in signal to  noise
ratio.
 
The baseband data were 2-bit digitised and the autocorrelation
functions (ACFs) computed by the correlator were folded at the
topocentric pulse period to form 90-s sub-integrations with 1024 bins
per lag. The ACFs were corrected for the non-linear effects of the
digitisation process and Fourier transformed to produce power spectra
for each sub-integration. Narrow-band interference  was rejected prior
to  dedispersion, the data were compressed to 32, 16 or 8 frequency
channels and saved to disk. A typical observation consisted of a
contiguous set of sixteen 90-s sub-integrations. Time synchronisation
was provided by starting the first sub-integration coincident with the
10 s pulse from the observatory's time system traceable to the NASA
DSN at Tidbinbilla using a microwave link, and from there to UTC(NIST)
via common-view Global Positioning System observations.

After summing the sub-integrations and frequency channels within each
observation, the highest signal to noise ratio (SNR) profiles were
added to form standard profiles for each observing frequency and
pulsar. The standard profiles at each frequency were cross-correlated
to determine their relative phase offset and  phase-shifted to
match. Pulse TOAs were determined  by cross-correlating the 24-min
integrated profiles with the corresponding standard
profile. Arrival-time data were fitted to a pulse timing model using
the {\small TEMPO} program \cite{tw89}. We used the DE200 ephemeris of
the  Jet Propulsion Laboratory \cite{sta82} to transform the TOAs to
the solar system barycenter and the Blandford \& Teukolsky (1976)
\nocite{bt76} model for timing pulsars in binary systems.

\setcounter{table}{0}
\begin{table*}
\begin{minipage}{160mm}
\caption{-- continued}
\begin{tabular}{lllll}
\hline\hline \\

Pulsar		     
&	PSR J1643$-$1224	  
&	PSR J1911$-$1114       	 
& 	PSR J2129$-$5721	  
&	PSR J2145$-$0750 	  
\\ \hline \\
		
R. A. (J2000)	     
&	16\th 43\tm 38\fs15583(4)  
&	19\th 11\tm 49\fs2927(2) 
&	21\th 29\tm 22\fs7539(3)  
&	21\th 45\tm 50\fs46813(4)  \\	

Decl. (J2000)        
&	$-$12\deg 24\min 58\farcs720(4)				  
&	$-$11\deg 14\min 22\farcs33(2)	
& 	$-$57\deg 21\min 14\farcs093(2) 		    	  
&	$-$07\deg 50\min 18\farcs361(2)				  \\
 	
PM in R. A. (mas y$^{-1}$)			  
&	3(1)			  
&	$-$6(4)
&	7(2)			  
&	$-$9.1(7)						  \\

PM in Decl. (mas y$^{-1}$)			  
&	$-$8(5)			  
&	$-$23(13)		
&    	$-$4(3)			  
&	$-$15(2)					          \\

Galactic longitude   
&	5\fdg67 		  
&	25\fdg14  		  
&	338\fdg01	 	  
&	47\fdg78		  \\ 

Galactic latitude    
&	21\fdg22   		  
&	$-$9\fdg58  		  
&    	$-$43\fdg57		  
&	$-$42\fdg08		  \\
  
Period (ms)	     
&	4.6216414465636(1)	  
&	3.6257455713977(5)	  
&	3.7263484187800(5)	  
&	16.0524236584091(2)       \\

Period derivative ($\times 10^{-20}$)  		  
&	1.849(1) 		  
& 	1.416(8)			
&	2.074(4)		  
&	2.986(1)						  \\

Epoch (MJD)	     
&	50288.0			  
&	50458.0 		  
&	50444.0		      	  
&	50317.0 		  \\

Dispersion Measure (cm$^{-3}$ pc)		  
& 	62.4121(2)		  
&	30.9750(9)	
&	31.855(4)		  
&	9.0006(1)		     				  \\

Orbital Period (days)				  
&	147.0173943(7) 		  
&	2.71655761(1)			
&  	6.62549308(5)		  
& 	6.838902510(1)						  \\

a ${\rm sin}i$ (light-s)			  
&	25.072613(1)		 
&	1.762875(5)
&	3.500559(3)		  
&	10.1641055(5)						  \\

Eccentricity	     
&	0.0005058(1)		  
&	0.0000188(55)		  
&	0.0000068(22)		  
&	0.0000193(1)		  \\

Epoch of periastron (MJD)			  
&	50313.03963(1000)	  
&	50455.59359(20000)		
&	50442.60321(20000)	  
&	50313.71221(1000)					  \\

Longitude of periastron (deg)			  
&	321.810(9)		  
&	178(22)
&	178(12)  		  
&	200.8(3)						  \\

Timing data span (MJD)				  
& 	49645$-$50932		  
&	49982$-$50933			
& 	49983$-$50931	 	  
&	49703$-$50933						  \\

RMS timing residual ($\mu$s)			  
&	5.5			  
&	1.4
&	1.2			  
&	1.4							  \\

Number of timing points			  	  
&	126			  
&	35
&	66			  
&	80							  \\
\hline
\end{tabular}
\end{minipage}
\end{table*}

\section{RESULTS}

Of the 17 MSPs discovered in the Parkes 70-cm survey, we have eight
binary and five isolated MSPs that show considerable improvement in
timing accuracy compared with results published in their respective
discovery papers \cite{bhl+94,lnl+95,llb+96,bjb+97}. The timing
parameters  resulting from the best fits to the TOAs are shown in
Table 1 where figures in parentheses represent 1$\sigma$ uncertainties
in the last significant digits  quoted. Table 1 shows we now have
finite values for the eccentricities of four MSPs where only upper
bounds existed previously. Finite values for the spin-down  rate of
two MSPs have also replaced previous bounds. Proper motions are listed
in Table 2. The nine pulsars  for which we have new arrival-time
proper motion measures are as follows; PSRs J0613$-$0200,
J1045$-$4509, J1643$-$1224, J1455$-$3330, J1603$-$7202, J1730$-$2304,
J1911$-$1114, J2129$-$5721 and J2145$-$0750. All but the first three
MSPs have  had their speeds determined in the scintillation study by
Johnston \etal (1998) \nocite{jnk98}.

Our timing analysis of PSR J1744$-$1134 has yielded a  measure of its
parallax. The parallax of $\pi$=2.7$\pm$0.4 mas corresponds   to a
distance of 370$^{+65}_{-47}$ pc (Toscano et al. in preparation),
significantly greater than the catalogued distance of 166 pc
\cite{tml93}. We have used this parallax-derived distance in our
calculations.

The proximity of PSR J1730$-$2304 to the ecliptic makes measurement of
its proper motion in declination difficult. Fitting for proper  motion
in ecliptic coordinates did not significantly improve  the proper
motion parameters. We have, however, used its motion in ecliptic
longitude as a lower bound to its total proper motion.
	
The average time spanned by our data was $\sim$1130 days with an
average of ninety 24 min observations per pulsar. Table 1 shows that
the typical post-fit rms was under 2 $\mu$s. The majority of  the fits
returned values of ${{\chi}^2}/{\nu}$ (where $\nu$ is the number of
degrees of freedom) of between 1.3 and 2.9. To compensate for
systematic effects we  multiplied  the TOA uncertainties by a factor
of between 1.1 and 2.7 chosen such that  ${{\chi}^2}/{\nu}$ was
unity. The trend in the TOAs for a small number of MSPs in our  sample
show low levels of unmodelled behaviour. The origin of these effects
is most likely a combination of low signal to noise integrated
profiles and the effects of radio frequency interference (RFI). At
higher frequencies our data do not show the same extent of RFI
contamination  and therefore are a better measure of the intrinsic
timing accuracy.

\section{DISCUSSION}

\subsection{Pulsar velocities}

\subsubsection{Pulsar velocity distributions}

The numerical modelling work of Tauris \& Bailes (1996) \nocite{tb96}
and Cordes \& Chernoff (1997) \nocite{cc97} and the standard models
for the formation of recycled pulsars \cite[and references
therein]{bv91}  all suggest MSPs have velocities much lower than their
long-period counterparts. 
These lower velocities are qualitatively understood in terms of the
asymmetry of the SN explosion and the fraction of mass suddenly ejected 
from the system during this event. These elements depend upon the 
characteristics of the system prior to the SN which, in turn depend upon 
a number of factors including; the rate and duration of mass loss 
and/or transfer, the nature of the common-envelope phase of binary evolution, 
and the peculiar evolution of each star.    
The distinction between the velocities of ordinary pulsars and
MSPs is evident from the histograms of pulsar velocities in Fig. 1.  A
Kolmogorov--Smirnov test shows that the probability that the
respective  velocities were drawn from the same distribution is
$<$10$^{-5}$.  Most of  the ordinary pulsars have velocities up to 500
km s$^{-1}$ with a small  number of pulsars with velocities as high as
$\sim$1000 km s$^{-1}$. The vast  majority of MSPs, however, have
velocities less than $\sim$130 km s$^{-1}$ and average just 85$\pm$13
km s$^{-1}$.  Assuming this velocity distribution is isotropic, the
rms total space velocity of MSPs is expected to be  $\langle
V^{2}\rangle^{1/2}$=($\frac{3}{2}V_{t}^{2}$)$^{1/2}$=129 km s$^{-1}$
with an rms velocity  in any one direction of   $\langle
V_{1}^{2}\rangle^{1/2}$=($\frac{1}{2}V_{t}^{2}$)$^{1/2}$=74 km
s$^{-1}$. Even after excluding proper motion measurements with values
less than 4$\sigma$, MSPs remain on average a factor  of four
slower. We present stronger evidence for the suggestion that  isolated
MSPs have lower velocities than binary MSPs \cite{jnk98}.  We note
that isolated MSPs are on average two-thirds slower than binary MSPs
and so any model for the formation and evolution of isolated MSPs
should take this into account.

The implications of such a comparison are, however, mitigated by
observational selection effects. Dispersion measure smearing and
scattering by the ISM of short-period pulsar signals reduces their
detectability. As a consequence the volume surveyed in a blind search
for pulsars will be considerably less for MSPs than for slower
pulsars.  Most of the observed MSPs are isotropically distributed
within $\sim$1 kpc of the Sun. Because the observed population of MSPs
is so `local' it is likely that high velocity pulsars moving
perpendicular to the plane of the Galaxy will have been selected
against in pulsar surveys.  Approximately two thirds of MSPs are found
in binaries whereas only $\sim$1 per cent of pulsars with periods
greater than 100 ms are in binary systems. The space velocity of
binary MSPs is expected to increase with decreasing orbital period
\cite{tb96}.  However, searches are less sensitive to short orbital
period pulsars because of orbital acceleration effects. Our data do
not show any significant correlation between velocity and orbital
period.   A comprehensive analysis  of the observational selection
effects is beyond  the scope of this paper, however, the 23 MSP
velocity measurements are qualitatively consistent with the
predictions of the aforementioned MSP models.

\setcounter{table}{0}
\begin{table*}
\begin{minipage}{180mm}
\caption{-- continued (Isolated MSPs)}
\begin{tabular}{llllll}
\hline\hline \\

Pulsar		     
&	PSR J0711$-$6830	  
& 	PSR J1024$-$0719     	 
&	PSR J1730$-$2304         
&    	PSR J1744$-$1134          
&    	PSR J2124$-$3358                  
\\ \hline \\

R. A. (J2000)	     
&	07\th 11\tm 54\fs21813(6)  
&	10\th 24\tm 38\fs7040(1) 
&	17\th 30\tm 21\fs6483(4) 
&	17\th 44\tm 29\fs390605(6) 
&	21\th 24\tm 43\fs86194(6)          \\	

Decl. (J2000)	     
&	$-$68\deg 30\min 47\farcs5793(4)				 

& 	$-$07\deg 19\min 18\farcs849(3)		
&  	$-$23\deg 04\min 31\farcs4(1)				 
& 	$-$11\deg 34\min 54\farcs5716(7) 			  
& 	$-$33\deg 58\min 44\farcs257(1)	  \\	

PM in R. A. (mas y$^{-1}$)			  
&	$-$15.7(5)		 
&	$-$41(2)	
&	20.5(4)			 
&	18.64(8)		 
&	$-$14(1)	          \\

PM in Decl. (mas y$^{-1}$)			  
&	15.3(6)			 
&	$-$70(3)
&    				 
&	$-$10.3(5)               
&	$-$47(1)                  \\

Galactic longitude   
&	279\fdg53 		  
&	251\fdg70	 	 
&	3\fdg14  		 
&	14\fdg79		  
&	10\fdg93		  	  \\ 

Galactic latitude    
&	$-$23\fdg28  		  
&	40\fdg52		 
&   	6\fdg02   		 
& 	 9\fdg18 		  
& 	$-$45\fdg44 		  	  \\ 

Period (ms)	     
&	5.4909684158061(1)	  
& 	5.1622045530114(7)	 
&	8.1227979128499(2) 	 
& 	4.07454587504741(2)	  
&	4.9311148591481(1)		  \\

Period derivative ($\times 10^{-20}$)		  
&	1.4900(1)		 
&	1.873(5) 			
& 	2.021(1)		 
&	0.8940(1)                
&	2.054(1) 		  \\

Epoch (MJD)	     
&	50425.0			  
&	50456.0			 
&	50320.0			 
&      	50331.0		          
& 	50288.0			  	  \\

Dispersion Measure (cm$^{-3}$ pc)		  
&	18.4106(3)		 
&	6.495(6)			
&	9.611(2)		
&	3.13775(4) 		 
&	4.6152(2)		  \\

Timing data span (MJD)				  
&	49922$-$50929		 
& 	49982$-$50930		 
&	49708$-$50931		 
&	49730$-$50932		 
&	49644$-$50933		  \\

RMS timing residual ($\mu$s)			  
&	2.1			 
&	4.6 
& 	2.0			 
&	0.67			 
&	5.2		          \\

Number of timing points			  	  
&	108			 
&	28 
& 	74			 
&	94			 
&	126		          \\
\hline
\end{tabular}
\end{minipage}
\end{table*}

\subsubsection{Pulsar dynamics and motion away from the plane}

In a study of the proper motions of 26 ordinary pulsars  Lyne,
Anderson \& Salter (1982) \nocite{las82} demonstrated that pulsars
were in general moving away from the Galactic plane; an observation
consistent with the view that pulsars are born in a population with a
scale height of 70 pc and move away from the plane with a z-velocity
of $\sim$100 km s$^{-1}$ \cite{go70}. The work of Harrison, Lyne \&
Anderson (1993) \nocite{hla93} brought the number of pulsars for which
the transverse velocity was not simply an upper limit to 66 and added
further weight to this argument. Harrison \etal also showed that the
small number of pulsars moving toward the plane must have been born at
large z-heights since they were not old enough to have undergone  at
least a quarter of a period of Galactic oscillation from a birthplace
near z$=$0. The lower panel of Fig. 2 shows the motion of the 89 ordinary 
pulsars with measured proper motions \cite{tml93} over a period of 1 Myr.
Although these plots do not take into account the effects of the
unknown  radial velocity, it is  evident that the general motion of
ordinary pulsars is directed away from the plane.

For a more accurate measure of the bulk motion of pulsars we chose to
eliminate pulsars with velocities less than 4$\sigma$. To avoid
including pulsars that were still within the progenitor layer ($|z|$
$<$ 200 pc) we have also eliminated pulsars with ($\tau_{c} \times v_{z}$)
$<$ 200 pc, where $\tau_{c}$ is the characteristic age in Myrs and $v_{z}$
is the  component of velocity perpendicular to the plane in km
s$^{-1}$. These  criteria leave us with a total of 24 pulsars. The
distribution of their Galactic  latitude velocity component is
symmetric about a mean of 24$\pm$55 km s$^{-1}$. Multiplying these
velocities by  $b/|b|$  assigns them a positive value if their motion
is away from the plane and negative otherwise. The distribution of
velocities multiplied by this factor has  a similar dispersion but
about a mean of 165$\pm$43 km s$^{-1}$. This implies that the majority
(20) are moving away from the plane.

The period of oscillation in the Galactic potential for a  pulsar on
the plane imparted with an initial z-velocity of 100 km s$^{-1}$ is
$\sim$150 Myrs \cite{oor65}. Although large compared with the  average
age of ordinary pulsars, this time is only a fraction of the typical
lifetimes  of MSPs ($\sim$1 Gyr). We would, therefore, expect MSPs to
form a dynamically old population that has reached dynamic 
equilibrium. Selection effects notwithstanding, we should observe as
many MSPs moving toward the plane as away. The upper panel of Fig. 2 
illustrates well the apparent isotropy of MSP motion over a period of 3 Myrs. 
Taking the 17 MSPs with velocities greater than 4$\sigma$ we find
their velocity distribution in $b$ has a mean of $-$27$\pm$16 km s$^{-1}$. 
Multiplying these velocities by $b/|b|$ results in a similar mean velocity 
of $-$20$\pm$17 km s$^{-1}$ suggesting that almost half the
pulsars were moving away from the plane.  This is in general
agreement with the velocity isotropy expected for such  an old
population.
 
Another clue to the age of MSPs is evident in their asymmetric
drift. A virialised population with a significant radial velocity
dispersion will rotate slower about the Galaxy than the local circular
speed \cite{mb81}.  Only an old population is expected to have come
close enough to a virialised state to show this effect. The signature
of asymmetric drift in this case would be an excess of  MSPs moving in
the opposite direction to the local flow. To see if this effect was
apparent we  corrected our proper motion derived  velocities for the
motion of the Sun with respect to the local standard  of rest. Because
of the unknown contribution of the radial velocity we removed from our
sample those MSPs which had $l$ within 25$^{\circ}$ of
$l$=90$^{\circ}$ or 270$^{\circ}$,  and $b$ within 25$^{\circ}$  of
the Galactic poles. We restricted ourselves to MSPs velocity
measurements more significant than 4$\sigma$  since  large errors can
reverse the apparent direction of the velocity.  This selection leaves
us with a total of ten MSPs of which seven are  moving in the
direction expected of asymmetric drift.  The average velocity of these
MSPs is 25$\pm$12 km s$^{-1}$ directed opposite to the local flow, a
clear manifestation of asymmetric drift and of  the dynamic
equilibrium of the MSP population.

\begin{figure*}
\setlength{\unitlength}{1cm}
\begin{picture}(11,10)
\put(-3,11){\includegraphics{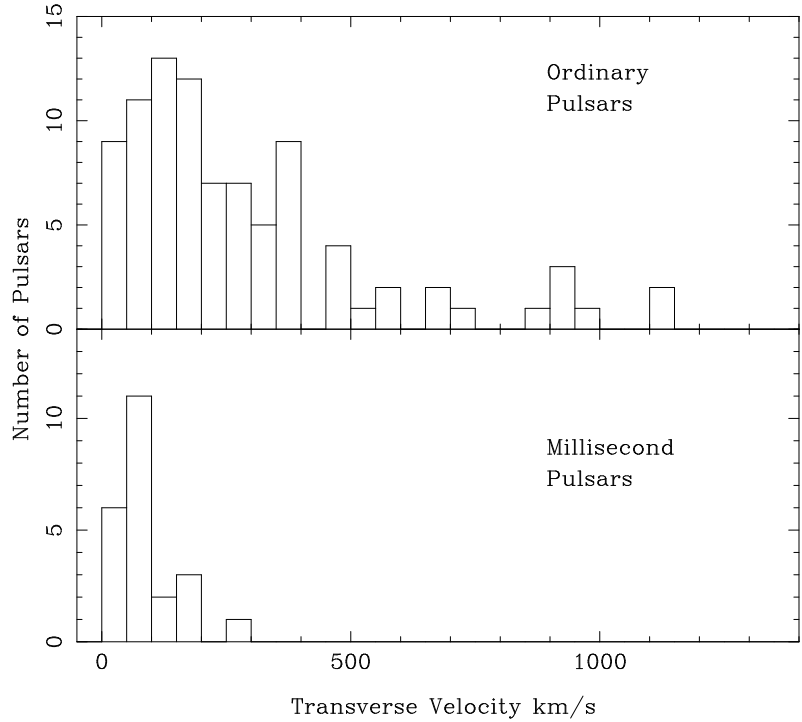}}
\end{picture}
\caption
[] {Histograms of transverse velocities for 89 ordinary pulsars
(P $>$ 20 ms) and 23 millisecond pulsars. The mean transverse velocity 
for MSPs is 85$\pm$13 km s$^{-1}$, approximately a factor of 
four slower than the mean transverse velocity of ordinary pulsars. 
A Kolmogorov--Smirnov test shows that the probability that these
velocities were drawn from the same distribution is $<$10$^{-5}$. }
\end{figure*}

\begin{figure*}
\setlength{\unitlength}{1cm}
\begin{picture}(5,10)
\put(-3,11){\includegraphics{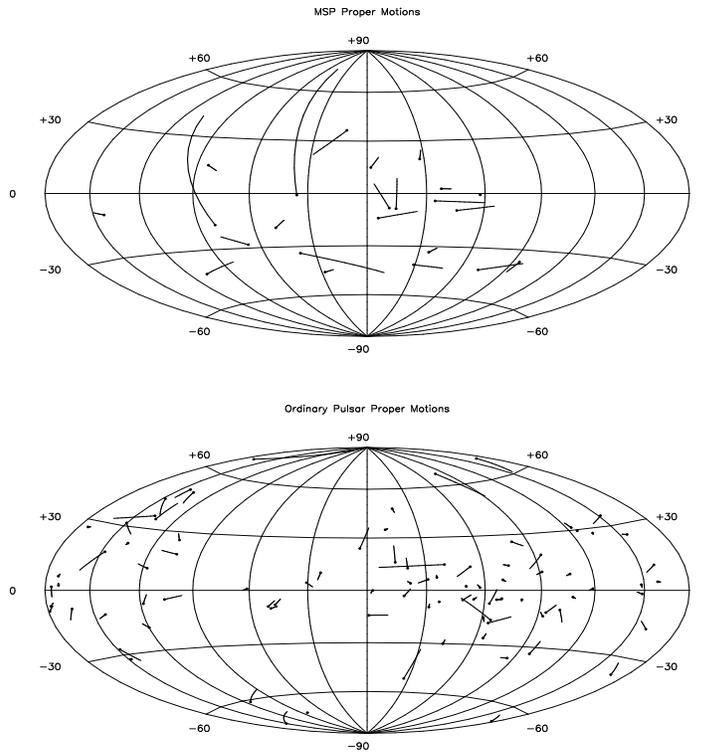}}
\end{picture}
\caption
[] {Aitoff--Hammer projections depicting the motion of ordinary and 
millisecond pulsars.  Tracks represent the predicted motion of pulsars over 
3 and 1 Myrs for MSPs and ordinary pulsars respectively (the final positions being 
marked by solid circles). The upper panel shows that almost as many MSPs are moving toward
the Galactic plane as away; in contrast to the motion of ordinary pulsars
(lower panel) which is in general directed away from the plane.}
\end{figure*}

\subsubsection{Comparison with scintillation velocities}

Of the 13 MSPs for which Johnston \etal (1998) \nocite{jnk98} have
measured scintillation speeds we have improved arrival-time proper
motions for four and new proper motions for six. The scintillation
speeds, $V_{\rm{ISS}}$, they present were derived using the following
equation \cite{gup95}:

\begin{equation}
V_{\rm{ISS}} = {3.85\times10^{4}} \left. \frac{(D_{\rm kpc} \Delta \nu_{\rm MHz} X)^{1/2}}
{f_{\rm GHz} t_{\rm iss,s}} {\rm km} \right. {\rm s}^{-1}.
\end{equation}

\noindent
Here $D$ is the distance to the pulsar, $f$ is the observing
frequency, $t_{\rm{iss}}$ and $\Delta\nu$ are respectively the
decorrelation time  and bandwidth, $X$ is the ratio of observer-screen
distance to screen-pulsar distance, and the constant
3.85$\times$10$^{4}$ is derived by  assuming a Kolmogorov turbulence
spectrum for the interstellar medium \cite{grl94}.

Johnston \etal found that for 9 MSPs, under the assumption that the
screen was  located  midway between the observer and the pulsar
(i.e. $X$=1), the average  ratio of the proper motion velocity to the
scintillation speed was 1.05. Using our new and improved proper motion
velocities we found that for 12 MSPs the average ratio was 0.98. Only
for 5 MSPs does the ratio differ from unity by more than  20 per cent:
0.7 for PSR J0437$-$4715, 0.8 for PSR J2129$-$5718, 0.6 for PSR
J2317+1439, 1.3 for PSR J0711$-$6830 and 1.4 for PSR
J2124$-$3358. This would suggest that the location of the scattering
screen is respectively 30, 36, 29, 64 and 67 per cent of  the way to
the pulsar. We have not included PSR J2051$-$0827 for which the
scintillation speed is more than a factor of three larger than the
proper motion velocity. It is likely that the wind from this MSP's
companion is effecting it's scintillation properties \cite{sbl+96} and
thereby contributing to the discrepancy in velocities.

\subsection{The intrinsic spin-down rates of MSPs}

\subsubsection{Background}

The observed spin-down rates for MSPs are approximately six orders of
magnitude lower than those of ordinary pulsars. As a consequence the
contribution to the measured period derivative, ${\dot{P}}_{m}$, from
acceleration effects, cannot be ignored.  Damour \& Taylor (1991)
\nocite{dt91} cite three Doppler  accelerations that may bias
${\dot{P}}_{m}$, namely; the effect of Galactic differential rotation
(GDR), vertical  acceleration in the Galactic potential, and the
transverse velocity  of the pulsar. We may express the intrinsic
spin-down rate as ${\dot{P}}_{i}={\dot{P}}_{m}-\delta\dot{P}$ where


\begin{eqnarray}
\delta\dot{P} = P\left[\frac{{-v_{0}}^{2}}{cR_{0}}\left(\cos l + 
\frac{(d/R_{0})-\cos l}{\sin^{2} l + [{(d/R_{0})-\cos l}]^{2}}\right)\cos b 
\right.  \nonumber \\
\left. + \frac{{v_{t}}^{2}}{cd} + \frac{a_{z}}{c} \sin b\right]. 
\end{eqnarray}

\noindent
The first term in this expression accounts for the effects of GDR with
the Sun's galactocentric radius and rotation velocity set to
$R_{0}=$8.0 kpc and $v_{0}=$220 km s$^{-1}$ respectively \cite{rei93}.
Here $l$ and $b$ are respectively the pulsar's Galactic longitude and
latitude. The second term in this equation was first identified by
Shklovskii (1970) \nocite{shk70} as the apparent acceleration caused
by the transverse velocity of the pulsar $v_{t}$ at a distance $d$,
while the last term corresponds to the vertical acceleration
experienced by the pulsar in the Galactic potential. The magnitude of
$a_{z}$ is derived from the Galactic potential model of Kuijken \&
Gilmore (1989) \nocite{kg89} and is given by

\begin{equation}
\frac{a_{z}}{c} = 1.08\times10^{-19}{\rm s}^{-1} 
\left[\frac{1.25z}{(z^{2}+0.0324)^{1/2}}+0.58z\right],     
\end{equation}

\noindent
where $z$=$d \sin b$ is the height above the Galactic disc in kpc.
In column 6 of Table 2 we list the derived $\delta\dot{P}$ for the 23
MSPs with known proper motions. The transverse velocities used were
derived from the proper motion measurements in column 2. For these
MSPs  the vertical acceleration term contributes the least to
$\delta\dot{P}$  with an average  value of $\sim$3$\times$10$^{-22}$,
while the  GDR effect had an average magnitude of
$\sim$5$\times$10$^{-22}$. By far the most significant effect comes
from the transverse velocity term which was on average
$\sim$9$\times$10$^{-21}$. Column 7 lists the intrinsic  period
derivatives, $\dot{P}_{i}$. In the majority of  cases these are
significantly less than the measured spin-down rates in column 5.

It is interesting to note that if PSR J1024$-$0719 is assumed to be
at its dispersion measure derived distance of 354 pc then the
magnitude of $\delta\dot{P}$ would be larger than that of its measured
period derivative. Since the $\delta\dot{P}$ correction of
2.9$\times$10$^{-20}$ is  dominated by the transverse velocity term
and the proper motion measure  is significant, we are certain that the
dispersion measure distance is an overestimate. Under the assumption
that the measured period derivative is completely  due to the
transverse velocity we can place a firm, model independent upper limit
on the distance to PSR J1024$-$0719 of 226 pc. A distance  to PSR
J1024$-$0719 of 200 pc, consistent with a parallax of $\pi$=5$\pm$4
mas (at the 1$\sigma$ level), and consistent with the aforementioned
upper  limit, was used in our calculations.

\begin{table*}
\begin{minipage}{160mm}
\caption{Derived quantities for MSPs with measured proper motions}
\begin{tabular}{lcccccccc}
\hline\hline \\

Pulsar & $\mu$ & Distance & Velocity & ${\dot{P}}_{m}$ & $\delta \dot{P}$ & ${\dot{P}}_{i}$ & $B_{i}$ & $\tau_{c,i}$ \\
    & (mas y$^{-1}$) & (kpc) & (km s$^{-1}$) & ($10^{-20}$) & ($10^{-20}$) & ($10^{-20}$) & ($10^{8}$ G) & (Gy) \\ \hline \\
J0437$-$4715 & 140.7(3) &     0.180   & 120.5(2) &     !!5.730  &     @4.932   &     @!!!0.7975  &     @3.0  &     @6.0  \\
J0613$-$0200 & 7(1)     &     2.190   & 77(11)   &     !!!0.9570 &    @!0.1207  &     @!!0.836  &      @1.6  &     @5.8  \\
J0711$-$6830 & 22.0(6)  &     1.039   & 78(2)    &     !!1.490  &     @!0.6557  &     @!!!0.8341 &     @2.2  &     @10   \\
J1024$-$0719 & 81(4)    &     0.200   & 62(3)    &     !!1.873  &     @1.574   &     @!!!0.2989 &      @1.3  &     @27   \\
J1045$-$4509 & 7(2)     &     3.246   & 52(15)   &     !!1.767  &     @!0.1312  &     @!!1.636  &      @3.5  &     @7.2  \\
B1257+12     & 95.0(8)  &     0.624   & 284(2)   &     11.43   &      @8.612   &     @!!2.821  &       @4.2  &     @3.5  \\
J1455$-$3330 & 24(13)   &     0.738   & 100(54)  &     !!2.423  &     @!0.9043  &     @!!1.519  &      @3.5  &     @8.3  \\
J1603$-$7202 & 8.6(7)   &     1.636   & 27(2)    &     !!1.571  &     @!0.4610  &     @!!1.110  &      @4.1  &     @21   \\
J1643$-$1224 & 8(5)     &     4.860   & 159(99)  &     !!1.847  &     @!0.8721  &     @!!!0.9751 &     @2.2  &     @7.5  \\
J1713+0747   & 6.4(2)   &     1.111   & 28.2(9)  &     !!!0.8536 &    @!0.1025  &     @!!!0.7511 &     @1.9  &     @9.6  \\
J1730$-$2304 & $>$20.2(4) &   0.509   & $>$51(1) &     !!2.021  &     !$>$0.4664  &  !!$<$1.554  &  $<$3.6  &  $>$8.3  \\
J1744$-$1134 & 21.3(4)  &     0.370   & 33.1(6)  &     !!!0.8941 &    @!0.1810  &     @!!!0.7131 &     @1.7  &     @9.1  \\
B1855+09     & 6.16(7)  &     1.000   & 17.3(2)  &     !!1.784  &     @!!0.05056 &     @!!1.733  &     @3.1  &     @4.9  \\
J1911$-$1114 & 24(13)   &     1.588   & 183(99)  &     !!1.415  &     @!0.8383  &     @!!!0.5764 &     @1.5  &     @10   \\
B1937+21     & 0.48(1)  &     3.580   & 79(2)    &     10.51   &      !$-$0.0467  &   @10.56  &        @4.1  &     @0.23 \\
B1957+20     & 30.4(8)  &     1.533   & 190(5)   &     !!1.685  &     @!0.5380  &     @!!1.147  &      @1.4  &     @2.2  \\
J2019+2425   & 23(2)    &     0.912   & 83(7)    &     !!!0.7022 &    @!0.4529  &     @!!!0.2491 &     @1.0  &     @25   \\
J2051$-$0827 & 5(3)     &     1.300   & 14(8)    &     !!1.272  &     @!!0.07892 &     @!!1.193  &     @2.3  &     @6.0  \\
J2124$-$3358 & 49(2)    &     0.247   & 53(2)    &     !!2.054  &     @!0.7547  &     @!!1.300  &      @2.6  &     @6.0  \\
J2129$-$5721 & 8(4)     &     2.550   & 56(28)   &     !!2.068  &     @!0.2569  &     @!!1.811  &      @2.6  &     @3.3  \\
J2145$-$0750 & 18(2)    &     0.500   & 38(4)    &     !!2.986  &     @!0.7794  &     @!!2.207  &      @6.0  &     @12   \\
J2317+1439   & 7(4)     &     1.892   & 94(53)   &     !!!0.2423 &    @!!0.09165 &     @!!!0.1504 &    @!0.73 &    @36   \\
J2322+2057   & 24(4)    &     0.782   & 80(13)   &     !!!0.9741 &    @!0.5504  &     @!!!0.4236 &     @1.4  &     @18   \\
\hline
\end{tabular}
\end{minipage}
\end{table*}

\begin{figure*}
\setlength{\unitlength}{1cm}
\begin{picture}(11,10)
\put(-3,11){\includegraphics{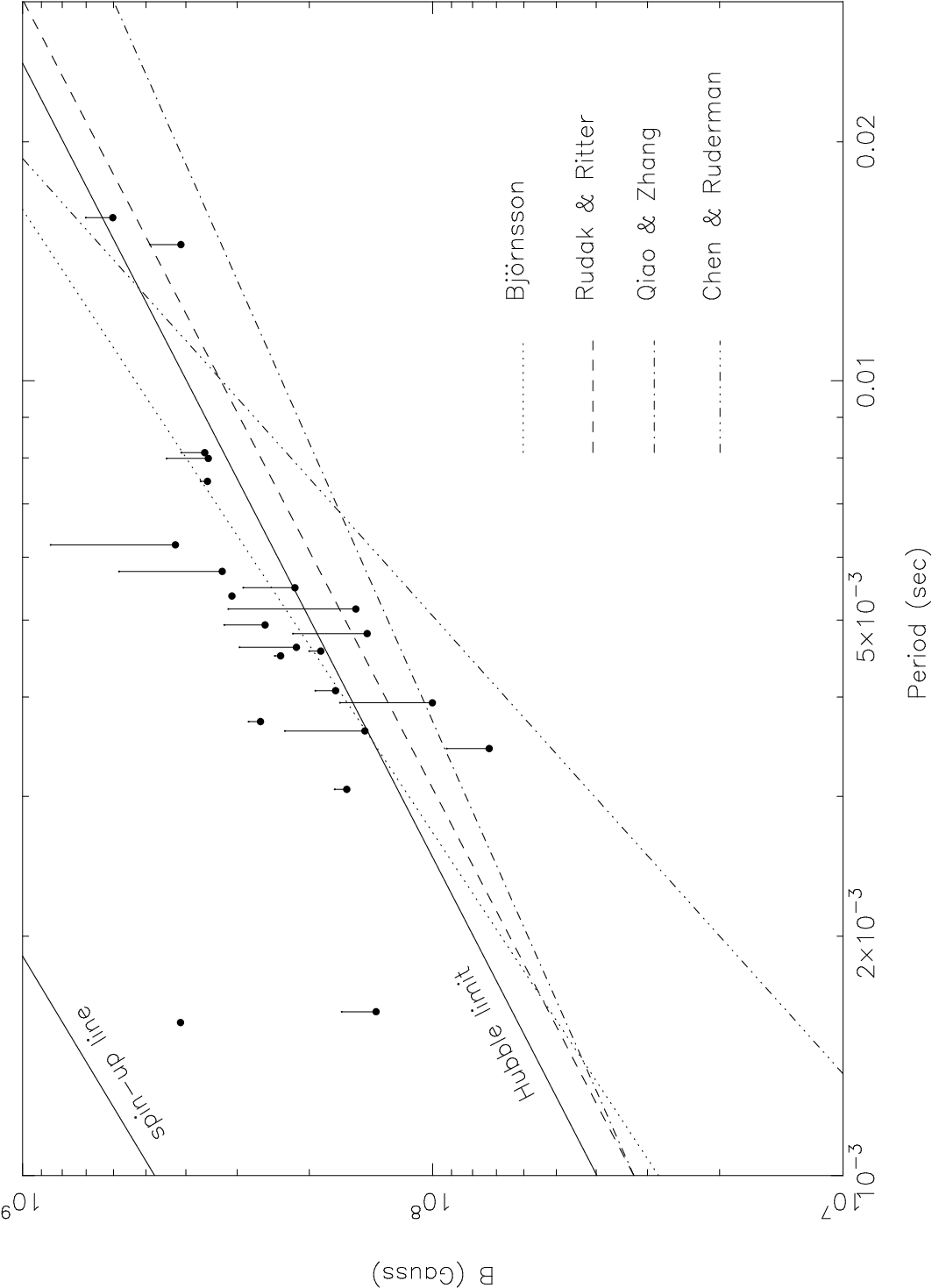}}
\end{picture}
\caption
[] {Magnetic field$-$Period diagram for the 23 MSPs with measured 
velocities. Tracks show the change in position of MSPs in the $B$--$P$ plane
when corrections are made to the spin-down rate to compensate for apparent Doppler 
accelerations. The positions corresponding to the intrinsic spin-down
rates are marked by solid circles. Also plotted are several death lines for 
MSPs, the Hubble limit (corresponding to an age of 10$^{10}$ years), and the
spin-up line (see text).}
\end{figure*}

\subsubsection{Magnetic fields, characteristic ages and death lines for MSPs}

Using a magnetic dipole spin-down model with a braking index $n$, the
age of a pulsar, $t$, may be related to its current period, initial
period, $P_{0}$, and spin-down rate thus,

\begin{equation}
t = \frac{P}{(n-1)\dot{P}} 
\left[ 1 - \left(\frac{P_{0}}{P}\right)^{n-1} \right].
\end{equation}

\noindent
Assuming $P_{0} \ll P$ and $n$=3 we obtain an expression for the
pulsar's `characteristic age' $t= \tau_{c} \approx P/(2\dot{P})$.  If the
surface magnetic field is taken to be dipolar then its magnitude is
given by $B \approx 3.2\times10^{19} (P\dot{P})^{1/2}$G, where $P$ is
in seconds. The last two columns of Table 2 list $B_{i}$ and
$\tau_{c,i}$, respectively the magnetic field strength and
characteristic ages calculated using the intrinsic spin-down rates.

In Fig. 3 we plot magnetic field versus rotation period for the MSPs,
showing the magnitude of the correction to the measured spin-down
rate. The correction applied not only results in a decrease in the
derived magnetic field strength but also an increase in characteristic
age.  Fig. 3 shows that, of the 23 MSPs, ten are  found to have ages
close to or greater  than 10 Gyr (the so-called `Hubble limit').  The
apparent paradox of MSPs with ages greater than that of the Galaxy
(9$\pm$2 Gyr; Winget \etal 1987\nocite{whl+87})  may be resolved in
several ways \cite{ctk94}.    Firstly, the assumption that $P_{0} \ll
P$ may be incorrect, and in fact, the `oldest' MSPs are born with
initial periods close to their current values. This would suggest that
the mass accretion rate in low-mass  systems is well below the
Eddington-limited rate. Secondly, the braking  index may be greater
than three because of multipolar magnetic field  structure or angular
momentum loss to gravitational radiation or indeed is time
variable. If MSPs do, however, possess true ages comparable to the age
of the Galaxy then their study would offer us the exciting prospect of
investigating the earliest processes of star formation in our Galaxy.

Most models of the spin-down evolution of pulsars predict a  cessation
of radio emission when electron-positron pairs can  no longer be
produced. The region of the $B$--$P$ space where the  emission
terminates for a particular pulsar depends upon the  structure and
magnitude of its magnetic field and defines the  so-called `death
valley' for  long-period pulsars \cite{cr93}.  The attempts to define
an equivalent `death line' for short-period  pulsars
\cite{rr94,qz96,bjo96}  have been plotted in Fig. 3. We note that no
single death line falls below  all the points.

\section{CONCLUSIONS}

In this paper we have presented improved timing parameters for 13 MSPs
including nine new proper motion measurements. Combining these results
with previous measurements brings the number of measured MSP
velocities to 23. The analysis of these velocities have lead to the
following conclusions:

\begin{itemize}

\item
The transverse velocity distribution of MSPs has a mean value of
85$\pm$13 km s$^{-1}$; approximately a factor of four slower than the
mean transverse velocity of ordinary pulsars. The MSP velocity
distribution we derived is qualitatively consistent with the
models of Tauris \& Bailes (1996) \nocite{tb96} and Cordes \& Chernoff
(1997) \nocite{cc97}, and the scintillation speeds measured by Johnston
\etal (1998) \nocite{jnk98}.

\item
MSPs form a dynamically old population. As expected for a population
that has reached dynamic equilibrium, almost as many MSPs are moving toward
the Galactic plane as away. This is in contrast with the young
population of ordinary pulsars which have a bulk motion directed away
from the plane. MSPs also show an asymmetric drift characteristic of a
virialised population.

\item
Our corrections to the spin-down rate of MSPs result in increases in
characteristic ages and decreases in the derived surface magnetic
field strengths.  In particular, almost half of the MSPs for which
this correction has been made have  characteristic ages  comparable
to, or greater than, the age of the Galaxy. Such large ages may be
accounted for if the birth periods are close to the  current periods.
		
\end{itemize}

\end{document}